# Dynamics of Magnetic Nano-Flake Vortices in Newtonian Fluids

Nasim Bazazzadeh[a], Seyed Majid Mohseni[a]*, Amin Khavasi[b], Mohammad Ismail Zibaii[c], S. M. S. Movahed[a], and Gholam Reza Jafari[a]

[a] Department of Physics, Shahid Beheshti University, Tehran, 19839, Iran
[b] Department of Electrical Engineering, Sharif University of Technology, Tehran, 11555-4363, Iran
[c] Laser and Plasma Research Institute, Shahid Beheshti University, Tehran, 19839, Iran

*corresponding author: m-mohseni@sbu.ac.ir, majidmohseni@gmail.com, Fax: +98 21 22431663, Phone: +98 21 2990 5044

Email addresses: n.bazazzadeh@gmail.com, majidmohseni@gmail.com, khavasi@sharif.edu, mizibaye@gmail.com, m.s.movahed@ipm.ir, gjafari@gmail.com

**Abstract:** We study the rotational motion of nano-flake ferromagnetic discs suspended in a Newtonian fluid, as a potential material owing the vortex-like magnetic configuration. Using analytical expressions for hydrodynamic, magnetic and Brownian torques, the stochastic angular momentum equation is determined in the dilute limit conditions under applied magnetic field. Results are compared against experimental ones and excellent agreement is observed. We also estimate the uncertainty in the orientation of the discs due to the Brownian torque when an external magnetic field aligns them. Interestingly, this uncertainty is roughly proportional to the ratio of thermal energy of fluid to the magnetic energy stored in the discs. Our approach can be implemented in many practical applications including biotechnology and multi-functional fluidics.

1. **Introduction**

Magnetic nanostructures and nanoparticles[1–3] have great potential in many nano-bio-magnetic applications, including bio-sensing[4–6], cell separation[7], magnetic resonance imaging contrast-enhancement agents[8], targeted drug delivery[2], deep brain stimulation[9], and targeted cancer-cell destruction[10,11]. Among them, magnetic nano-flakes, with thickness below 100 nm and approximate radii well below a micron, a circularly-closed magnetzation profile, nominated as magnetic vortex, has shown great interest in multidisciplinary science including magnetism[12] and bio-technology[11]. The formation of topologically robust magnetization with in-plane component except with that having out-of-plane counterpart at the centre represents negligible stray field. This behaviour provides a different type of weak interacting particles analogue to that of superparamagnetic particles. Applying in-plane magnetic field can displace the vortex centre and results in large in-plane magnetization[12]. This leads to strong magnetic response of nano-flake vortices (NFVs) which gives them an advantage over superparamagnetic nanoparticles[13]. They show interesting properties, e.g. zero remanence and strong magnetic response. Albeit magnetic vortex is a deeply focused and important subject in spintronics[14], its particle-like response in fluids for different application purposes is still open. For those vortices which are freely dispersed in liquid, by applying arbitrary magnetic field, any changes in magnetization can be transferred to a mechanical torque to rotate them in parallel orientation to the magnetic field direction[11,13]. This was a major impact that NFVs used to destruct the glioblastoma multiforme (GBM) tumor cell by applying alternative magnetic field and damaging the cell body by mechanical forces[11]. In addition, such vortices with about tens of pico-Newton force can be in many bio-logical systems to stimulate the cell and provide forces inside the cell by applying magnetic field far from the specimen, analogue to a very new system suggested in [9]. Therefore, in most of applications, understanding Brownian dynamics of such vortices to optimize their impact for a proposed work, e.g. here maximizing the force, is of great importance.

The simulation and direct visualization of rotational Brownian motion of ellipsoid disc-like particles has been reported in[15]. However, this work only considers two dimensional motion with no external magnetic field. There are also some experimental works indirectly investigating the rotational Brownian motion of NFVs under applied magnetic field[11,13]. The Brownian motion of different types of particles suspended in a Newtonian fluid has been numerically studied[16–21]. More recently, the numerical simulation of stochastic rotation of tri-axial ellipsoidal magnetic particles under constant and time-varying magnetic fields has been presented in[22]. Nevertheless, there is a lack of rigorous theoretical studies for dynamics of NFVs in fluids. Also an analytical approach is needed to better understand the phenomenon.

In this work, we derive the stochastic equation for NFVs in the dilute limit conditions, taking into account the effects of shear flow and the special magnetic torque. To find the dynamical response of NFVs, the equation is then solved in the time domain with alternating applied magnetic field. We also propose an approximate analytical solution for this equation. The maximum error of the proposed analytical model will be studied which is correlated with misalignment of NFVs against applied magnetic field due to their Brownian rotation. We present a physical picture for describing this misalignment based on the thermal and magnetic energies. Finally, we investigate the accuracy of both numerical and analytical simulations by comparing them against experimental results.

2. **Rotational Brownian dynamics for NFVs**

We consider a disc of radius $R$ and thickness $L$ ($L<<R$) with arbitrary orientation, and will find out the mechanical rotation in the present of a uniform external magnetic field **H**. The magnetic field directed along the $z$-axis of the laboratory reference frame (unprimed axes), as shown in Fig. 1.

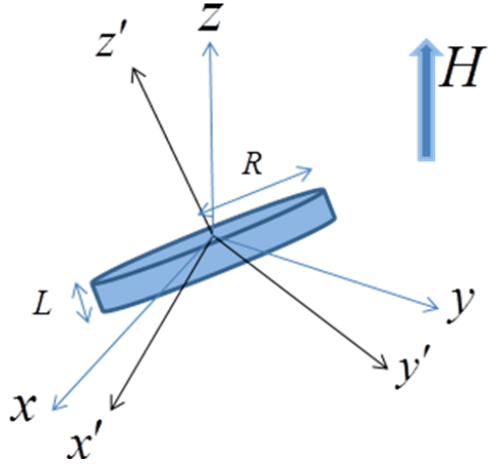

Fig. 1. A disc with an arbitrary orientation with respect to the applied magnetic field. The primed and unprimed axes show the particle and laboratory reference frames, respectively.

Let us define a new reference frame (primed axes) whose $z$-axis is along the disc axis. Writing equations in the particle primed axes is more straightforward because all the torques are exerting on the particle and the equations can be simply expressed in the particle coordinate. Toward this, we have to only transform the applied magnetic field from the laboratory reference frame to that of the particle. The transformation can be carried out by taking the Euler angles into account. On the other hand, we pursue to study the evolution of Euler angles against external parameters.

Now, there are twelve possible rotation sequences where we choose the rotation sequence $z$-$x$-$z$ (these rotations are shown in Fig. 2(a), (b) and (c), respectively). Because, the magnetic field is along the $z$ axis and the angle between the disc axis ($z'$ axis) and the applied magnetic field is what we are looking for. In this sequence just one rotation is not around the $z$ axis and thus the $z$ axis rotates only once as shown in Fig. 2(b). As a result, the angle between the $z$ and $z'$ axes is simply $\theta$ as shown in Fig. 2 (c).

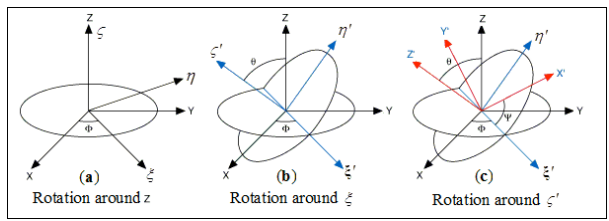

Fig. 2. Transformation between primed and unprimed axes with the rotation sequence $z$-$x$-$z$: (a) rotation around the $z$ axis by $\phi$, (b) rotation around the $\xi$ axis (rotated $x$ axis) by $\theta$ and (c) rotation around the $\zeta'$ axis (rotated $z$ axis) by $\psi$.

The vectors in the primed frame are related to the unprimed one by

$$\mathbf{H'} = \mathbf{AH} \qquad (1)$$

where

$$\mathbf{A} = R_\psi R_\theta R_\phi \qquad (2)$$

is the transformation matrix and it is given in terms of rotation matrixes $R_\phi$, $R_\theta$ and $R_\psi$, which are representing rotations around the $z$, $\xi$ (rotated $x$) and $\zeta$ (rotated $z$) axes, respectively (these rotations are depicted in Fig. 2(a), (b) and (c)). The rotation matrices can be easily obtained from the following equations

$$R_\varphi = \begin{pmatrix} \cos(\phi) & \sin(\phi) & 0 \\ -\sin(\phi) & \cos(\phi) & 0 \\ 0 & 0 & 1 \end{pmatrix} \qquad (3)$$

$$R_\theta = \begin{pmatrix} 1 & 0 & 0 \\ 0 & -\sin(\theta) & \cos(\theta) \\ 0 & \cos(\theta) & \sin(\theta) \end{pmatrix} \qquad (4)$$

$$R_\psi = \begin{pmatrix} \cos(\psi) & \sin(\psi) & 0 \\ -\sin(\psi) & \cos(\psi) & 0 \\ 0 & 0 & 1 \end{pmatrix} \qquad (5)$$

Now, we can write the angular momentum equation. In the primed frame, the stochastic angular momentum equation under Stokes-flow conditions and for distance between NFVs, r, with $r<<R(Re)^{-1}$ where $Re$ is the Reynolds number, can be written as [23]

$$\boldsymbol{\tau}'_h + \boldsymbol{\tau}'_m + \boldsymbol{\tau}'_B = 0 \qquad (6)$$

where $\boldsymbol{\tau}'_h$, $\boldsymbol{\tau}'_m$ and $\boldsymbol{\tau}'_B$, are hydrodynamic, magnetic and Brownian torques, respectively. The hydrodynamic torque for a disk is

$$\boldsymbol{\tau}'_h = -R_{T\omega}\boldsymbol{\omega}' \qquad (7)$$

where $\boldsymbol{\omega}'$ is the rotational velocity of the particle. In this equation, for a disc-shaped particle $R_{T\omega} = 32\eta_0 R^3/3$ denotes the rotational hydrodynamic resistance where $\eta_0$ is the shear viscosity. For the sake of simplicity, we exclude the Brownian torque in this step. It will be included at the final step when we write the equation in terms of Euler angles. Therefore, equation (6) is simplified to

$$\boldsymbol{\omega}' = R_{T\omega}^{-1}\boldsymbol{\tau}'_m \qquad (8)$$

The only unknown parameter in (8) is the magnetic torque which can be obtained using the following equation

$$\boldsymbol{\tau}'_m = \int_{disc} \mu_0 \mathbf{M}' \times \mathbf{H}' dv' = \mu_0 \left( \int_{disc} \mathbf{M}' dv' \right) \times \mathbf{H}' \qquad (9)$$

where $\mathbf{H}'$ is the applied magnetic field in the primed coordinate, $\mu_0$ is the vacuum permeability and $\mathbf{M}'$ is the magnetization inside of the NFV. If the applied field is much smaller than the saturation field, the average magnetization will be a linear function of the in-plane component of the applied field

$$\langle \mathbf{M}' \rangle_V = \chi(0) \mathbf{H}'_\parallel \qquad (10)$$

where $\chi(0)$ is the initial susceptibility which can be calculated analytically [12],[24], and $\mathbf{H}'_\parallel = \mathbf{H}' - H'_z \hat{\mathbf{z}}'$. Now, substituting (10) in (9) and after some straightforward mathematical manipulations, we obtain

$$\boldsymbol{\tau}'_m = \mu_0 V \chi(0) \mathbf{H}'_\parallel \times H'_z \hat{\mathbf{z}} \qquad (11)$$

where $V$ is the NFV volume.

The rotational dynamics of the NFV is described by substituting (11) in (8), in the particle coordinate system. However, the main purpose of this work is to study of how the primed frame changes in time with respect to the laboratory frame. In other words, we are interested in dynamics of Euler angles. So, we have to obtain a relation between the Euler angles and the angular velocity in the primed frame ($\boldsymbol{\omega}'$). The idea is to consider small variations in each Euler angle, and determine the effects on the rotation vector. The first Euler angle experiences two additional rotations, namely $R_\psi R_\theta$, the second angle one rotation, namely $R_\psi$, and the final Euler angle no additional rotations[25]

$$\boldsymbol{\omega}' = R_\psi R_\theta \begin{pmatrix} 0 \\ 0 \\ \frac{d\varphi}{dt} \end{pmatrix} + R_\psi \begin{pmatrix} \frac{d\theta}{dt} \\ 0 \\ 0 \end{pmatrix} + \begin{pmatrix} 0 \\ 0 \\ \frac{d\psi}{dt} \end{pmatrix} \qquad (12)$$

Hence, by substituting (3), (4) and (5) in (12) we have

$$\begin{pmatrix} \omega'_x \\ \omega'_y \\ \omega'_z \end{pmatrix} = \begin{pmatrix} \sin\psi \sin\theta & \cos\psi & 0 \\ \cos\psi \sin\theta & -\sin\psi & 0 \\ \cos\theta & 0 & 1 \end{pmatrix} \begin{pmatrix} \frac{d\varphi}{dt} \\ \frac{d\theta}{dt} \\ \frac{d\psi}{dt} \end{pmatrix} \qquad (13)$$

where $\omega'_x$, $\omega'_y$ and $\omega'_z$ are the x, y and z components of the vector $\boldsymbol{\omega}'$, respectively.

Now, let us include the Brownian torque effect exerted on an NFV which is suspended in a fluid. Each component of the Brownian term ($R_{T\omega}^{-1} \boldsymbol{\tau}_B$) can be modeled by a random white noise, $\eta(t)$, with zero mean and variance of $\langle \eta(t)\eta(t') \rangle = 2k_B T R_{T\omega}^{-1} \delta(t-t')$ where $k_B$ is the Boltzmann constant [15]. This leads to a random rotation angle $\Delta \boldsymbol{\Theta}'_B$. The vector elements of $\Delta \boldsymbol{\Theta}'_B$ are modelled by a Gaussian distribution with zero mean and variance of $\langle \Delta\theta'_B \Delta\theta'_B \rangle = 2k_B T R_{T\omega}^{-1} \Delta t$ [26]. Here, we can include the effect of the Brownian rotation angle to (13) as follows

$$\begin{pmatrix} \Delta\varphi \\ \Delta\theta \\ \Delta\psi \end{pmatrix} = \begin{pmatrix} \sin\psi \csc\theta & \cos\psi \csc\theta & 0 \\ \cos\psi & -\sin\psi & 0 \\ -\sin\psi \cot\theta & -\cos\psi \cot\theta & 1 \end{pmatrix} \begin{pmatrix} \omega'_x \\ \omega'_y \\ \omega'_z \end{pmatrix} \Delta t + \Delta \boldsymbol{\Theta}'_B \quad (14)$$

This equation can be numerically solved resulting in the time dependence of Euler angles.

3. **Approximate analytical model**

In this section, we derive an analytical solution for (14). To this end, we divide the problem into two parts where the applied magnetic field could be I) nonzero and II) zero. Implementing the first condition leads to a nonlinear Langevin equation. To solve this equation we neglect the stochastic term, namely the Brownian torque. We in addition estimate the maximum error determined by neglecting this term which is small enough in most realistic cases. However, for the second situation we derive an exact analytical solution.

*3-1. Strong magnetic field*

When a fairly strong external magnetic field is present, it exerts a considerable torque on NFVs. Although the Brownian term is large and comparable with the magnetic torque in some cases, it should be noted that its average is zero. Therefore, when we study the cumulative effect of a large number of NFVs (e.g. for the light transmitted through a solution of NFVs [11,13]), the Brownian term may be negligible.

It should be also noted that the NFV is symmetric with respect to its axis. Hence, rotations around this axis, $R_\phi$ and $R_\psi$, has no effect on the problem. Considering $\psi = \phi = 0$ and assuming the applied magnetic field as $\mathbf{H} = H\hat{\mathbf{z}}$, from (1) we have

$$\begin{pmatrix} H'_x \\ H'_y \\ H'_z \end{pmatrix} = \begin{pmatrix} 1 & 0 & 0 \\ 0 & \cos\theta & \sin\theta \\ 0 & -\sin\theta & \cos\theta \end{pmatrix} \begin{pmatrix} 0 \\ 0 \\ H \end{pmatrix} \quad (15)$$

Substituting this equation in (11) and then in (8), the following relation can be obtained

$$\boldsymbol{\omega}' = R_{T\omega}^{-1} \mu_0 V \chi(0) H^2 \sin(\theta)\cos(\theta) \hat{\mathbf{x}} \quad (16)$$

Finally, from (14) and (16) with $\psi = 0$ the following differential equation is derived

$$\frac{d\theta}{dt} = \beta \sin(2\theta) \qquad (17)$$

where $\beta = R_{T\omega}^{-1} \mu_0 V \chi(0) H^2 / 2$. This equation can be analytically solved

$$\theta(t) = \tan^{-1}\left[\tan\theta_0 \exp(2\int_{t_0}^{t} \beta dt)\right] \qquad (18)$$

where $\theta_0 = \theta(t_0)$ is the initial orientation of the NFV.

Let us estimate the maximum error of this solution. Our approximation becomes worse as the magnetic torque, $\beta \sin(2\theta)$, reduces. For a given $\beta$, the worst case (i. e. the maximum error) occurred for $\theta = 0$ and $\pi/2$. However, the case $\theta = 0$ is unlikely because all NFVs must be perpendicular to the applied field; thus we examine $\theta = \pi/2$. Let's assume that there is a small deviation in NFVs angle with respect to the magnetic field: $\bar{\theta} = \pi/2 - \theta$, due to the Brownian torque. So, we can rewrite (17) as

$$\frac{d\bar{\theta}}{dt} \approx -2\beta\bar{\theta} + \eta(t) \qquad (19)$$

in which the effect of the Brownian torque, $\eta(t)$, is included and $\beta \sin(2\theta)$ is approximated with $-2\beta\bar{\theta}$. Equation (19) is the well known linear Langevin equation. For sufficiently long times (still much shorter than field cycling) the average of $\bar{\theta}$ is zero and its variance is given by[27]

$$\langle \bar{\theta}^2 \rangle = \frac{2k_B R_{T\omega}^{-1} T}{4\beta} = \frac{k_B T}{\mu_0 V \chi(0) H^2} \qquad (20)$$

Roughly speaking, the maximum error in calculating $\theta$ is proportional to the ratio of thermal energy of fluid to the magnetic energy stored in NFVs (assuming $\chi(0) \approx 1 + \chi(0)$ or $\chi(0) >> 1$).

*3-2. Zero magnetic field*

When no external magnetic field exists, the first term in (14) is zero because there is no magnetic torque. In this situation, we have a random walk problem. As already mentioned, the steps of random walk model have Gaussian distribution with zero mean and variance of $\sigma^2 = 2k_B T R_{T\omega}^{-1} \Delta t$. So, for each Euler angle we should solve a random walk problem which step length, $x$, has the following probability density function (PDF)

$$P(x) = \frac{1}{\sqrt{2\pi}} \exp\left(\frac{-x^2}{2\sigma^2}\right) \qquad (21)$$

In order to calculate the PDF of the distance from the origin after $N$ steps, we first define the characteristic function which is the inverse Fourier transform of the PDF

$$Z_x(k) = \int_{-\infty}^{\infty} \exp(ikx) P(x) dx = \exp\left(\frac{-k^2\sigma^2}{2}\right) \qquad (22)$$

After $N$ steps of random walk, the characteristic function will be $Z_x^N(k)$ [28]. Taking the Fourier transform of $Z_x^N(k)$, the PDF for the distance from the origin after $N$ steps, $s$, is

$$P_N(s) = \frac{1}{\sqrt{2\pi N \sigma^2}} \exp\left(\frac{-s^2}{2N\sigma^2}\right) \qquad (23)$$

As a result, the time-dependent PDF for the angle between the NFV axis and the z axis, $\theta$, can be simply computed as

$$P(\theta, t) = \frac{1}{\sqrt{4\pi k_B T R_{T\omega}^{-1} t}} \exp\left(\frac{-(\theta-\theta_0)^2}{4 k_B T R_{T\omega}^{-1} t}\right) \qquad (24)$$

where $t = N\Delta t$.

## 4. Results and discussion

### 4.1. Comparison with experimental data

In this section, we demonstrate the accuracy of the proposed numerical and analytical models by comparing their results with those of experimental report in[11]. The experiment has been carried out for $Ni_{80}Fe_{20}$ NFVs with $R$ = 500 nm and $L$ = 60 nm suspended in an aqueous solution (we assume the shear viscosity of water for this solution, i.e. $\eta_0 \approx 9 \times 10^{-4}$ Pa.s at 300 K). A periodic time-varying magnetic field, alternating between 0 and 10 Oe, has been applied to the solution including randomly dispersed NFVs in water. An optical beam parallel to the applied field transmits through the solution. The optical transmission is modulated by NFVs orientation: when the field is turned on, the plane of NFVs is aligned to the field direction and, as a result, the optical transmission reaches to its maximum value. On the other hand, when the applied field is turned off the randomness of NFVs is gradually recovered due to the Brownian rotation and thus the optical transmission decreases.

We carry out the simulation with $10^5$ NFVs and estimate that the intensity of the transmitted light, $I$, is proportional to $\langle |\sin\theta| \rangle$. The initial susceptibility of NFVs is assumed to be $\chi(0)$ = 20, taken from the experimental data in[24].

The simulated optical transmission for an applied field with frequency $f$ = 1 Hz, is plotted in Fig. 3 (a). When the field is on, the NFVs become parallel to the applied field and the optical transmission is high. Contrarily, when the field is off, NFVs will have random orientation. It can be seen that the numerical and analytical models virtually overlap. The rise time and fall time of the light intensity are 4.46 ms and 165 ms, respectively which is in good agreement with those experimental values (~5 ms and ~150 ms) reported in[11]. The rise of light intensity, which is due to the NFVs alignment with the applied field, is magnified in Fig. 3 (b). Moreover, the maximum error of analytical model is shown in the inset which is around 0.3%.

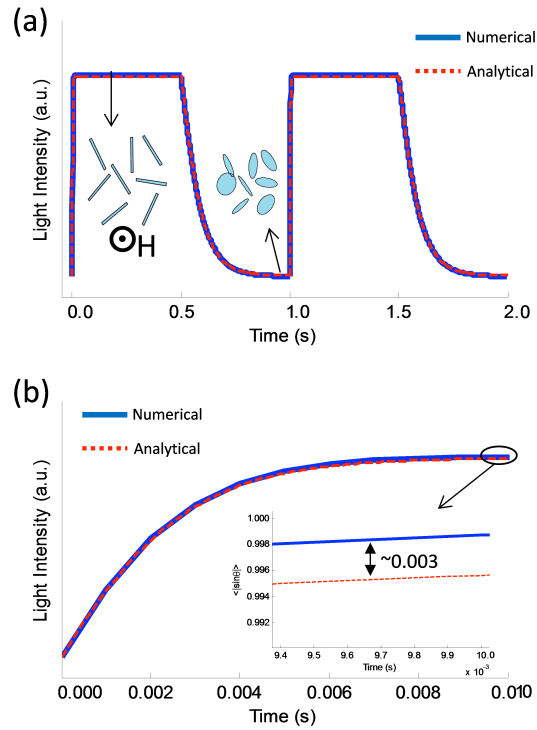

Fig. 3. (a) The intensity of the transmitted light through an aqueous solution of NFVs driven by a 1Hz uniform magnetic field, computed by the proposed numerical and analytical models. The NFVs orientations for "on" and "off" field are also illustrated in the inset. (b) The rise of light intensity when the magnetic field is switched on. Inset in panel (b): the maximum error of analytical model in the estimation of light intensity which is around 0.3%.

To further demonstrate the accuracy of the proposed models, the light intensity modulation $\Delta I = I_{max} - I_{min}$ as a function of frequency normalized to the maximum light intensity modulation ($\Delta I$ when $f \to 0$), are compared against experimental data (extracted from Fig. 2c of [11]) in Fig. 4. Again an excellent agreement is observed.

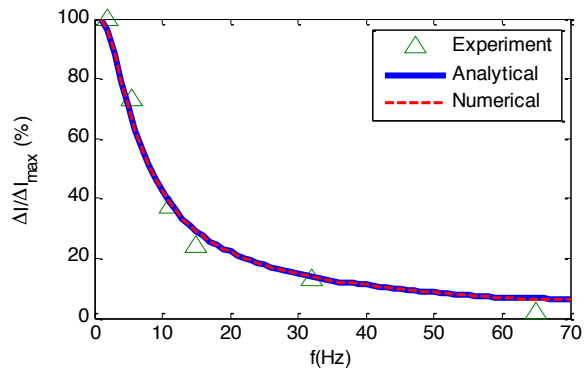

Fig. 4. The light intensity modulation in terms of frequency.

*4.2. Limitation of the model*

It is worth mentioning that the accuracy of the analytical model decreases when the rotational motion of only one NFV is considered. This is obvious in Fig. 5 where the rotational motion of one NFV is simulated by the numerical and analytical models. The difference between the two approaches is due to the Brownian term where supposed to be neglected in the analytical model. However, this term is of random nature with zero mean, so the error is negligible when considering a large number of NFVs, especially for large applied magnetic fields. It should be emphasized that the proposed analytical model is exact in the absence of external magnetic field.

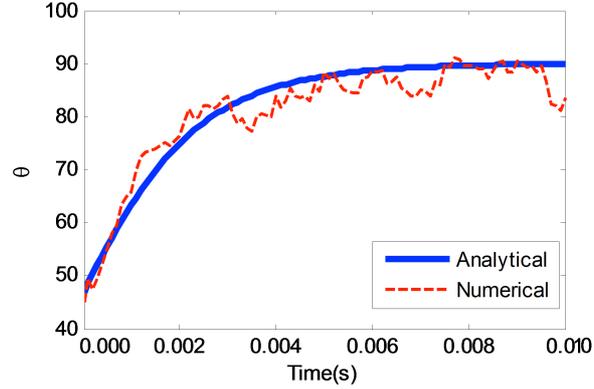

Fig. 5. The evolution of the Euler angle $\theta$ for one NFV in a uniform magnetic field. The initial value for $\theta$ is set to 45°.

### 4.3. Controllability of the NFVs

Let us examine the maximum error of the proposed model in calculating the transmitted light intensity ($\langle|\sin\theta|\rangle$) in the presence of the magnetic field. Inasmuch as we neglect the effect of the Brownian term in our analytical model, we obtain $\langle|\sin\theta|\rangle = 1$ from (18) as $t \to \infty$. However, we can analytically estimate the error of the proposed analytical model by using (20) in the $t \to \infty$ limit. To this end, we assume a Gaussian distribution for $\theta$ with the mean $\pi/2$ and the variance given by (20). So we can easily estimate $1-\langle|\sin\theta|\rangle$ as $t \to \infty$ by this argument. This parameter is plotted in Fig. 6 versus the applied magnetic field. The results are also verified by rigorous numerical simulations. It can be seen from this figure that for small values of magnetic field the orientation of NFVs is completely random leading to the constant value of 0.3633 for $1-\langle|\sin\theta|\rangle$. This value is below 0.01 for $|\mathbf{H}| > 5.3$ Oe. These results are of importance from practical point of view, because one can predict the minimum amount of the magnetic field required for controlling NFVs and how much the orientation of NFVs deviates from the desired direction due to the Brownian effect.

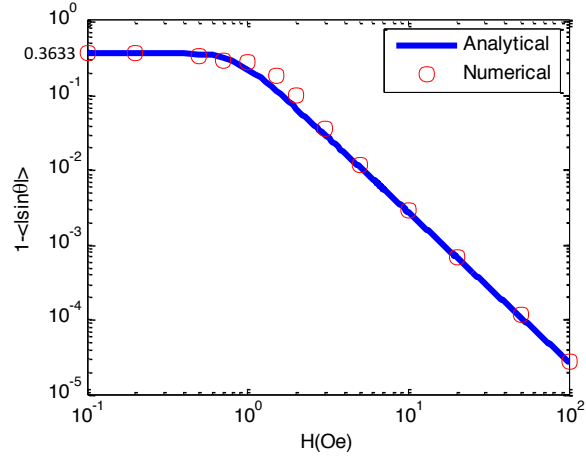

Fig. 6. Deviation from maximum light transmission ($1-\langle|\sin\theta|\rangle$) as a function of the applied magnetic field. The analytical curve is obtained using the approximate relation given in (20), while the numerical results are rigorous. This deviation is due to the Brownian rotation of NFVs which are misaligned with the applied magnetic field.

We also predict the effect of NFVs size on their response to the applied magnetic field. We calculate the minimum required field to ensure that $1-\langle|\sin\theta|\rangle < 0.01$. This is plotted in Fig. 7 (a) and (b) as a function of NFV thickness and radius, respectively. It is seen that larger the NFVs radius, smaller the required field. However, the dependence of the required field on the NFVs thickness is not strong and it slightly changes around 5 Oe, with a minimum around $L = 20$ nm.

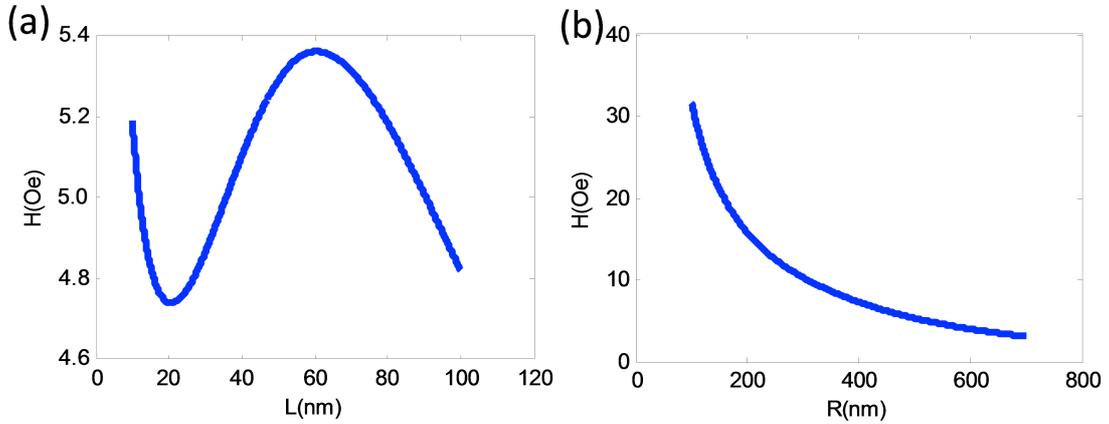

Fig. 7. The external field required for controlling the direction of NFVs (i.e. $1-\langle|\sin\theta|\rangle < 0.01$) versus the disc (a) thickness (with $R = 500$ nm) and (b) radius (with $L = 60$ nm). It is seen that there is a thickness around $L = 20$ nm where the required field is minimized. However, in terms of disc radius the required field is a monotonically decreasing function.

## 5. Conclusions

Although the Brownian dynamics of different types of suspended particles has already been studied[16–19,22,29,30], there is a lack of theoretical study on magnetic NFVs. In this work, we presented Brownian dynamics study of magnetic NFVs suspended in a Newtonian fluid, under an external time-varying magnetic field in the dilute limit. The creation of magnetic vortex state inside the discs leads to strong interaction of the discs with the applied field and therefore provides mechanical torque to such randomly oriented NFVs. We included in our model the hydrodynamic torque, the Brownian torque and the special magnetic torque which stems from the interaction of the vortex state with the external field. Not only the derived equation can be solved numerically, but also we proposed an analytical approach for solving this equation. At first, we examined the proposed model validation with those available experimental data for light transmission through liquid including NFVs[11]. We found that our approach can explain the dynamical responses of NFVs very well. In addition, thanks to the proposed analytical model, we predicted the magneto-mechanical susceptibility of NFVs against the amplitude of magnetic field. We also derived an analytical expression which predicts the deviation of NFVs with respect to the applied field due to the Brownian rotation. The effect of NFVs size on the required field to provide the rotation was also investigated.

This work leads to a better understanding of the rotational behavior of magnetic NFVs and might be of great importance in the investigation of biomedical applications related to magnetic-vortex nano-flakes especially in cell stimulation where mechanical effect is important[11]. This work is also helpful for future study of magnetically modulated optical transmission by magnetic nanoparticles suspended in fluids[13] which can be used in realizing magnetically tunable colloidal micromirrors[31]. The magnetoviscosity of dilute suspensions[19,20,30] of NFVs can be also investigated in the future works.